\documentclass[12pt]{article}
\usepackage{amsmath,amssymb,graphicx}
\usepackage{amsthm}
\usepackage{euscript}

\newtheorem{lem}{Lemma}
\newtheorem{thm}{Theorem}

\newcommand{\pr}{\noindent{\bf Proof}. }
\newcommand{\rem}{\noindent{\bf Remark}. }

\newcommand{\pa}{\partial}
\newcommand{\one}{\cO(1)}

\newcommand{\const}{\textrm{const}}

\newcommand{\hs}{ \hspace{1cm}}

\newcommand{\B}{\Big}
\newcommand{\blan}{\Big  \langle} 
\newcommand{\bran}{\Big  \rangle}

\newcommand{\ud}{\underline{d}}  

\newcommand{\be}{\begin{equation}}
\newcommand{\ee}{\end{equation}}
\newcommand{\bs}{\begin{split}}
\newcommand{\es}{\end{split}}

\newcommand{\al}{\alpha}
\newcommand{\De}{\Delta}
\newcommand{\de}{\delta}

\newcommand{\Ga}{\Gamma}

\newcommand{\Om}{\Omega}

\newcommand{\si}{\sigma}

\newcommand{\cC}{{\cal C}}
\newcommand{\cD}{{\cal D}}

\newcommand{\cO}{{\cal O}}

\newcommand{\cH}{{\cal H}}

\newcommand{\cF}{{\cal F}}
\newcommand{\cK}{{\cal K}}

\newcommand{\cL}{{\cal L}}

\newcommand{\cY}{{\cal Y}}

\newcommand{\bbR}{{\mathbb{R}}}
\newcommand{\bbZ}{{\mathbb{Z}}}
\newcommand{\bbC}{{\mathbb{C}}}

\begin{document}

\title{A Feynman-Kac  formula for magnetic monopoles}
\author{ 
J. Dimock
\thanks{dimock@buffalo.edu}\\
Dept. of Mathematics \\
SUNY at Buffalo \\
Buffalo, NY 14260 }
\maketitle

\begin{abstract} We consider the quantum mechanics of a charged particle in the presence of Dirac's magnetic monopole.
Wave functions are sections of a complex line bundle and the magnetic potential is a connection on the bundle.  
We use a continuum eigenfunction expansion to find an  invariant domain of essential 
self-adjointness for the Hamiltonian.  This leads to a proof of the  a Feynman-Kac formula  expressing solutions of the imaginary time Schr\"{o}dinger equation as stochastic integrals. 
   \end{abstract}

\section{Introduction}  \label{bundle1} 

Consider a single charged particle (an electron) in the presence of   a magnetic potential  one-form $A= \sum_k A_k dx^k$  with magnetic field 2-form  
$B = dA $.
The quantum mechanical Hamiltonian for this  particle is 
an operator $H$ on $L^2(\bbR^3)$ which has the form 
\be
H  = -  \frac12   \sum_{k}( \pa_k  - i A_k(x)  )^2 
\ee
The Feynman-Kac formula expresses the  semi-group $e^{-tH}$ in terms of stochastic integrals.   If $X_t$ is Brownian motion in $\bbR^3$
starting starting at $x$,  
then the equation is for $f \in L^2(\bbR^3)$
\be
(e^{-tH} f )(x)  = E^x\B(  \exp\B( - i \int_0^t   \sum_k A_k(X_t)  \circ dX^k_t  \B) f(X_t) \B)
\ee
Here $\int_0^t   \sum_k A_k(X_t)  \circ dX^k_t  $    is the Stratonovich stochastic integral.
This has been rather  thoroughly studied, often in the presence of an additional scalar potential.   See for example  Hinz  \cite{Hin15}  for references.

In this paper we establish a version of this for Dirac's magnetic monopole. 
The situation is  more difficult since a  magnetic monopole  creates a magnetic field   for which there is no smooth magnetic potential.
Instead one   has to formulate the problem on a Hilbert space of  sections of a certain  $U(1)$ line bundle  over $M= \bbR^3 - \{ 0\}$,  and replace
the potential by a connection on this bundle.

Our treatment is quite explicit, and uses a continuum eigenfunction expansion for the Hamiltonian to identify a domain of essential self - adjointness which is invariant under the associated semi-group. 
This makes it possible to give an efficient proof of a Feynman-Kac
formula.   General  treatments of the  Feynman-Kac formula for  vector bundles  on  a manifold have been given by   Norris  \cite{Nor92} and G\"{u}neysu \cite{Gun10}, 
\cite{Gun12}.

In section \ref{bundle2}  we define the bundle and the connection.  In section \ref{bundle3} we  introduce the   eigenfunction expansion
and use  it   to define the Hamiltonian as
a self-adjoint operator.  In section \ref{bundle4} we review some facts about stochastic integrals.  In section \ref{bundle5}  we define stochastic parallel
translation.  Finally in  section \ref{bundle6} we prove the Feynman-Kac formula.

\section{The monopole bundle}  \label{bundle2} 

The $U(1)$  bundle  $E$ is a manifold together with a smooth map $\pi: E \to M$ such that the fibers $E_x= \pi^{-1} x$
are vector spaces  isomorphic to $\bbC$.  
We take an  covering of $M = \bbR^3 - \{0\}$   by two open set $U_{\pm} $.
In   spherical coordinates     they are defined by 
 \be
 \begin{split}
U_+ = & \B \{ x \in M: 0 \leq  \theta  <  \frac{\pi}{2} +  \al \B\}    \\
U_-  = & \B \{ x \in M:  \frac12 \pi -  \al <  \theta    \leq \pi \B\}   \\
\end{split}
\ee
Here   $0 < \al  <  \frac12 \pi$  is a fixed angle. If $\de = \sin \al $ then  $0 < \de < 1$ and $\cos (\frac{\pi}{2}  \pm \al )  = \mp \de$.
Then in Cartesian coordinates 
 \be
 \begin{split}
U_+ = & \B \{ x \in M:  1 \geq   \frac{x_3}{|x| }  >  - \de  \B\}    \\
U_-  = & \B \{ x \in M:  \de   >  \frac{x_3}{|x| }   \geq  -1 \B\} \\
\end{split}
\ee
$E$ is defined so  in each region there is a trivialization which is a diffeomorphism
\be
h_{\pm} :  \pi^{-1} (U_{\pm}  )  \to  U^{\pm}  \times \bbC
\ee
such that for $x \in U_{\pm} $ the map $h_{\pm} :E_x \to \{ x \} \times \bbC$ is a linear isomorphism.  
These maps are related  in $U_+ \cap U_-$  by the transition functions for fixed integer $n$
\be
h_{+ }  h_{-}^{-1}   =   e^{2 i  n \phi }  
\ee
which means that if $v \in E_x$ and $h_{\pm} v = (x, v_{\pm} )$ then  $v_+ =  e^{ i 2 n \phi(x) }v_-$.
Concretely $E$ can be constructed   as equivalence classes  in $ M \times \bbC$ with  $(x, v_+) \sim (x, v_-)$ if  $x \in U_+ \cap U_{-}$
and $v^+ =  e^{ i 2 n \phi(x) }v_-$.    There is an inner product on $E_x$ defined unambiguously by $<v,w> = \overline{v_{\pm}} w_{\pm}$, and there is a norm  $|v| = |v_{\pm}|$.

A connection can be   defined by a one- forms  $A^{\pm} $  on $U_{\pm}$ which  in $U_+ \cap U_{-} $ are related by the gauge transformation 
\be \label{gauge}
A^{+ } = A^{-} +   2n d \phi 
\ee
They are defined by  
\be   \label{walnut} 
A^{\pm}  = - n( \cos \theta \mp 1 ) d \phi   =  n \B( \frac{ x_3}{|x| } \mp 1 \B) \frac {x_2 dx_1- x_1 dx_2}{x_1^2 + x_2^2}   
\ee
These each  yield the magnetic field $B$  of a  monopole of strength $n \in \bbZ, n \neq 0$: 
\be  \label{acorn1}
dA^{\pm} =n \sin \theta\ d \theta \wedge d \phi    =  \star \frac{n}{r^2} dr  \equiv  B    
\ee
Neither magnetic  potential  $A^{\pm}$ is singular,   but there is no one-form $A$ defined on all of $M$ such that $dA = B$, which is 
why we need the vector bundle.    The transition functions were designed to compensate  the gauge transformation (\ref{gauge}).

Let us check that $A^{+ }$  on $  U_+$ is not singular and get an estimate on its size.    It has a possible  singularity  on the $x_3$ axis  from the factor $(x_1^2+x_2^2)^{-1} $.  
However we can write  
\be
\frac{x_3}{|x|} -1 =   \frac{1}{|x|} (x_3-|x| )   = -   \frac{1}{|x|} \frac {x_1^2 + x_2^2 }{ x_3 + |x|} 
\ee
and hence
\be   \label{walnut2} 
A^+   =    \frac{-   n }{|x| ( x_3 + |x|) } 
 ( x_2 dx_1- x_1 dx_2 )    
\ee
Since $x_3 + |x|$ is bounded away from zero  this is smooth  on $U_+$.   
Indeed   $x_3 \geq -\de |x| $ on $U_+$,   hence  $x_3 + |x| \geq (1- \de ) |x| $, 
and  hence $x_i(x_3 + |x|)^{-1}  \leq (1- \de) ^{-1} $.   Thus for all   $x \in U_+ $ we have
\be \label{itsy}
|A^+_k(x)| \leq  |n|  (1-\de)^{-1} |x|^{-1} 
\ee

Parallel translation on a curve $x: [0,t] \to M$  is a linear isomorphism $\Pi_t: E_{x_0} \to E_{x_t}$ defined as follows. 
Let $v \in E_{x_0}$ and suppose the curve is  entirely  contained in some   $  U_{\pm} $.   If $h_{\pm} v = (x_0,  v_{\pm} )$ then 
$\Pi_t v$ is defined by 
\be \label{gang} 
h_{\pm} ( \Pi_t v)  =     (x_t, (\Pi_t v) _{\pm} ) =    (x_t, \Pi_t^{\pm} v_{\pm} )
\ee
where   $\Pi_t^{\pm}$ is multiplication by  (summation convention) 
\be
 \Pi_t^{\pm}=   \exp \B(  i    \int_0^t  A^{\pm} _k(x_s)  dx^k_s \B) 
\ee
 If  the curve is entirely contained in both $U_{\pm} $ then (\ref{gauge}) implies $  \Pi_t^{+} = e^{2in\phi(x_t) }  \Pi_t^{-} e^{-2in\phi(x_0)} $. Together with 
  $v_+ =  e^{2in\phi(x_0)} v_-$ this implies that  $\Pi_t^{+} v_{+} = e^{2in\phi(x_t) } \Pi_t^{-} v_-$ and hence the definition of $\Pi_t$ in
  (\ref{gang})  is independent of the choice $U_{\pm} $. 
Finally 
parallel translation  for a curve not contained in a single trivialization  can be defined by patching together pieces which stay within one trivialization.

 Now we  can define  a covariant derivative on sections of $E$.  A section of $E$ is a map $f : M \to E$ such that $\pi (f (x) )= x$  which says  $f(x) \in E_x$.  The set of 
 all smooth sections is denoted $\Ga(E) $.    For  $x \in M$ and
 $e_k$ the standard basis for $\bbR^3$ consider   parallel transport  along the curve $ x_t = x + t e_k$.     For  $f \in \Ga(E) $ define  $\nabla_k f \in \Ga(E) $
as the limit in $ E_{x}$  
\be
(\nabla_k f ) (x)  =  \lim_{t \to 0}  t^{-1} \B(   \Pi_t^{-1}  f( x_t ) - f(x) \B)
\ee   
If   $x \in U_{\pm} $  and $h_{\pm} f(x)  = (x, f_{\pm} (x) )$   this is computed as
\be
(  \nabla_k f ) _{\pm} (x)    =  \B( (\pa_k  - i  A^{\pm} _k ) f_{\pm}  \B)   (x) 
\ee

\section{The  Hamiltonian} \label{bundle3}

\subsection{definitions}  \label{defn} 

The Hamiltonian for our problem   is initially defined on smooth sections   $\psi \in \Ga(E)$  
by
\be
 H\psi  = - \frac12  \B( \sum_k \nabla_k \nabla_k   \B) \psi
\ee
We want to define it as a self-adjoint operator on the Hilbert space $\cH  = L^2(E,dx)$ consisting of all measurable
sections $\psi$  such that  $ \| \psi\|^2 = \int_M  | \psi (x) |^3 dx < \infty$. 

We use the strategy of Wu-Yang
 \cite{WuYa76} and the author \cite{Dim20}.  
First change to spherical  coordinates.  For any function $\Psi$  on  $M = \bbR^3-\{ 0\}$ we define $\hat \Psi $ on $\bbR^+ \times S^2$
by 
\be
\hat \Psi(r, \theta, \phi) = \Psi( r \sin \theta  \cos \phi,   r \sin \theta  \sin  \phi, r \cos \theta)
\ee
If $\Psi$ is a section of $\pi: E \to M$,  then $\hat \Psi$ is a section of a vector bundle  $\pi:  E'  \to \bbR^+ \times S^2$. 
With  $U_{\pm}' = U_\pm \cap S^2$ the bundle  $ E'$ has trivializations on $R^+ \times U' _{\pm}$  
still  with   transition functions $h_+h_-^{-1} =e^{2in \phi }$.  
For any $\Psi  \in  \cC^2(E) $ we have $\hat \Psi \in \cC^2(\hat E) $ and $(H\Psi)^{\wedge}   = \hat H \hat \Psi $
where 
\be \label{sunny2} 
\hat H f= \frac12\B(  - \frac{1}{r^2} \frac{\pa} {\pa r} r^2  \frac{\pa f} {\pa r} +  \frac{1}{r^2} ( \cL^2 - n^2)  \B) 
\ee
Here $\cL^2=  \cL_1^2+ \cL_2^2 + \cL_3^2$   and  $\cL_i$  are  angular momentum operators.    In the trivialization on  $\bbR^+ \times U'_{\pm} $ we have 
  \be
\begin{split}
\cL^{\pm} _1  =&   i  \B( \sin \phi  \frac{\pa}{ \pa \theta } + \cot \theta \cos \phi \frac{\pa}{ \pa \phi  } \B)  - n \cos \phi \left( \frac{1 \mp  \cos \theta}{\sin \theta}  \right)  \\
\cL^{\pm} _2  =&   i  \B( - \cos \phi  \frac{\pa}{ \pa \theta } + \cot \theta \sin  \phi \frac{\pa}{ \pa \phi  } \B)  - n \sin \phi \left( \frac{1 \mp  \cos \theta}{\sin \theta}  \right)  \\
\cL^{\pm} _3 = &  - i  \frac{\pa}{ \pa \phi  } \mp n  \\
\end{split}
\ee

The  map $\Psi \to \hat \Psi$ is unitary from $\cH =L^2(E,dx) $ to $\hat \cH =   L^2(E' , r^2dr d \Om ) $
where $d \Omega = \sin \theta d\theta d \phi$ is the Haar measure on $S^2$.  
In fact since the
the transition functions only depend on the angular variable $\phi$ we can make the identification 
\be 
\hat \cH =     L^2(\bbR^+ , r^2 dr )  \otimes L^2(\tilde E, d \Om )  
\ee
where $\tilde E$ is a vector bundle $\pi: \tilde E \to  S^2$   with trivializations on $ U'_{\pm}$   which still satisfy $h_+ h_-^{-1} = e^{2in \phi} $.
Now in  (\ref{sunny2})   the $\cL^2- n^2$ only acts on the factor  $L^2(\tilde E, d \Om)$.

The  joint spectrum of  the commuting operators $\cL^2, \cL_3$ has been studied by Wu-Yang \cite{WuYa76}. 
   They find a complete set of eigenfunctions $\cY_{n,\ell,m}(\theta, \phi )$
which are sections of $L^2(\tilde E,  d \Om  )$. They take  values $\ell  \geq |n|  $ and $|m| \leq \ell$ and satisfy
\be
\cL^2 \cY_{n,\ell,m} = \ell(\ell+1)   \cY_{n,\ell,m}  \hs  \cL_3  \cY_{n,\ell,m} =  m    \cY_{n,\ell,m}
 \ee
 The explicit expression for $\cY_{n, \ell, m} $ is given in section \ref{bubbles}.
Let  $\cK_{n,  \ell} $ be the eigenspace spanned by $\cY_{n,\ell,m}$ with $|m| \leq \ell$.
Then the Hilbert space can be identified with
\be
\hat  \cH  = \bigoplus_{\ell = |n| } ^{\infty}    L^2(\bbR^+, r^2dr)  \otimes  \cK_{n,\ell}  
\ee 
The Hamiltonian is now 
 \be
\hat H = \bigoplus_{\ell = |n|}^{\infty}    h_{\ell}   \otimes I 
\ee
where on smooth functions
 \be
h_{\ell} = \frac12  \B( - \frac{d^2}{dr^2} - \frac{2}{r}\frac{d}{dr}   +   \frac{\ell(\ell+1)  -n^2}{r^2}     \B)   
\ee
\bigskip

 \subsection{radial eigenfunctions}   \label{bingo} 
 
 To further study the radial  Hamiltonian $h_{\ell}$ 
we make a continuum eigenfunction
expansion.
Continuum eigenfunctions for $h_{\ell} $ are
\be
  (kr)^{-\frac12}  J_{\mu} (kr)   \hs  \mu^2 =   \ell(\ell + 1)  -n^2 + \frac14  
 \ee
 where  $J_{\mu} $ is the Bessel function of order $\mu>0$ regular at the origin.   We have 
   \be \label{pingpong} 
 h_{  \ell} \B(  (kr)^{-\frac12}   J_{\mu} (kr)  \B )  = \frac12  k^2  \B (  (kr)^{-\frac12}   J_{\mu} (kr) \B  )     
   \ee
For future reference we note that  since $\ell \geq |n| \geq 1$ we
have
 \be
 \mu  \geq \frac12 \sqrt{5} \geq 1.12
\ee

 The eigenfunction expansion is  the Fourier-Bessel transform and we recall some relevant facts  \cite{Tit48}, \cite{Dim20}.   The transform is defined by the formula
\be  \label{pinstripe} 
 \psi^{\#}_{\mu}  (k)= \int_0^{ \infty}    (kr)^{-\frac12}   J_{\mu}(kr)     \psi (r) r^2  dr
  \ee
at first  for say $\psi \in \cC_0^{\infty} (\bbR^+ ) $. 
It satisfies 
\be
 \int_0^{ \infty}   | \psi^{\#}_{\mu}  (k) |^2 k^2 dk  = \int_0^{ \infty}     |  \psi (r) |^2 r^2  dr
\ee
and  
extends to a unitary  operator  from
$L^2( \bbR^+, r^2 dr)$ to $L^2( \bbR^+, k^2 dk) $.
The inverse is given by exactly  the same formula
\be
 \label{pulchritude}
 \psi (r)= \int_0^{ \infty}    (kr)^{-\frac12}   J_{\mu}(kr)     \psi^{\#} _{\mu}  (k) k^2  dk
  \ee

 Now define the dense domain 
 \be
 \cD_0  =\{ \psi \in L^2(\bbR^+ , r^2dr):   \psi^{\#}  \in    \cC_0^{\infty} (\bbR^+ )   \}
  \ee
To analyze this domain we use the fact that 
$ x^{-\frac12} J_{\mu} (x) $ is a bounded smooth function with the asymptotics 
 \be \label{sappy} 
 x^{-\frac12} J_{\mu} (x)   =  \begin{cases} \cO( |x| ^{ \mu - \frac12} ) &   x \to 0   \\  \cO( |x|^{-1} )  &   x \to \infty \\
 \end{cases} 
 \ee

  \begin{lem}   \label{pinsky} 
    $ \cD_0 \subset \cC^2(\bbR^+ ) $.  If $\psi \in \cD_0$
    we have the asymptotics for $m=0,1,2$
    \be \label{slober2} 
\psi^{(m)}  (r)   =
 \begin{cases}  \cO(r^{\mu- \frac12 -m } ) & r \to 0 \\  \cO(r^{-1-m}  ) & r \to  \infty \\
\end{cases} 
\ee
Furthermore for $\psi \in \cD_0$ we have     $\psi, \psi', \psi'' \in L^2(\bbR^+ , r^2dr )$.
\end{lem} 
\bigskip

\rem  In fact  $\cD_0  \subset \cC^{\infty} (\bbR^+) $, and if we worked harder using Bessel function identities  as in \cite{Dim20},  we could show  the long distance asymptotics are  $\cO(r^{-N} ) $
for any $N$. 
\bigskip
 
 \pr $ \psi^{\#}  \in    \cC_0^{\infty} (\bbR^+ )  $ means that $k$ is bounded away from zero and infinity.   So the  asymptotics  for $\psi$    follow from (\ref{sappy}).   We
 know $\psi \in    L^2(\bbR^+ , r^2dr )$.

For the derivative  differentiate (\ref{pulchritude}) under the integral sign and then integrate by parts to obtain 
\be
 \begin{split}   \psi '(r)
 = &  \int_0^{ \infty}   d/dr   \B( (kr)^{-\frac12}  J_{\mu }(kr)\B)
 \psi^{\#}   (k)  k^2 dk  \\
 = &  \int_0^{ \infty}    \frac{k}{r}  \frac{d}{dk}    \B(   (kr)^{-\frac12}  J_{\mu}(kr) \B)  
 \psi^{\#}   (k)  k^2 dk  \\
 = & \frac{-1}{r}  \int_0^{ \infty}    (kr)^{-\frac12}   J_{\mu}(kr)   \B(3 \psi^{\#}   (k) + k(\psi^{\#} )'   (k) \B) k^2 dk   \\
\end{split}  
\ee
So  $\psi'(r) $ is $r^{-1}$ times a function in  $\cD_0$ and the asymptotics  follow.
Further $\psi' \in L^2(\bbR^+ , r^2dr )$ since  for $r\geq 1$ it is bounded by a function in $\cD_0 \subset L^2$,
and for $r \leq 1$ we have
\be
\int_{r \leq  1} | \psi'(r) |^2 r^2 dr  < \one \int_{r \leq  1}( r^{\mu - \frac32} )^2  r^2   dr   < \one  \int_{r \leq  1} r^{2\mu - 1}   dr  < \infty
\ee 

Similarly  $\psi''(r) $ is $r^{-2} $ times a function in  $\cD_0$ and the asymptotics follow. 
For   $\psi'' \in L^2$   the relevant integral is  $\int_{r \leq  1} r^{2\mu - 3}   dr  < \infty$.  (Higher derivatives would not be in $L^2$) . 
This completes the proof.

\begin{lem} { \ }  \label{sanpedro} 
 \begin{enumerate}
 \item  
For   $\psi \in \cD_0$ we have $h_{\ell} \psi \in \cD_0$ with 
$( h_{\ell} \psi )^\#(k)  = \frac12 k^2 \psi^{\#} (k)$
\item 
$h_{\ell}$ is self-adjoint on $D(h_{\ell})  = \{ \psi \in L^2(\bbR^+, r^2dr):  \frac12k^2  \psi^\# \in L^2(\bbR^+, k^2dk)\}$ 
and   essentially self-adjoint on $\cD_0$.
\item   
For $\psi    \in L^2(\bbR^+, r^2dr)$ we have $ ( e^{-h_{\ell} t}  \psi )^\#(k)  = e^{-\frac12 k^2t}  \psi^{\#} (k) $.  Hence if
  $\psi \in \cD_0$  then  $e^{-h_{\ell} t}  \in \cD_0$  
\end{enumerate}
 \end{lem}

 \pr \begin{enumerate} 
 \item In (\ref{pulchritude}) differentiate under the integral sign   and use (\ref{pingpong}) 
 \item  
 multiplication by $\frac12k^2$ is self-adjoint on  
   on $\{ f \in L^2 :  \frac12k^2 f \in L^2\} $ and essentially self-adjoint
   on $\cC^{\infty}_0(\bbR^+)$.   The stated result is the unitary transform of these facts.  
\item The unitary   transform $\psi \to \psi^\#$ provides the spectral resolution  $h_{\ell} $
and hence the definition of  $e^{-h_{\ell} t} $.  
 \end{enumerate}

\subsection{angular eigenfunctions} \label{bubbles} 

 We  consider  the monopole harmonics $\cY_{n,\ell,m} $ as defined by Wu-Yang  \cite{WuYa76}.  They are given  in  the trivializations on  $U'_{\pm} = U_{\pm} \cap S^2 $ by 
 \be \label{elf1} 
 \cY^{\pm} _{n, l,m}(\xi , \phi )   =  \const (1- \xi)^{\frac12 \al    } (1+ \xi  )^{\frac12 \beta} P^{\al  ,\beta}_{ \ell + m } (\xi )  e^{i(m \pm n) \phi }   \hs \xi = \cos \theta 
 \ee
 where   $\al =  -n-m, \beta = n-m $   and $P^{\al  ,\beta}_{ \ell + m }$ are Jacobi polynomials given    by 
\be \label{elf2} 
P^{\al  ,\beta}_{ \ell + m } (\xi) =  \const (1- \xi)^{- \al    } (1+ \xi )^{ - \beta} \frac{d^{\ell + m}} { d \xi^{\ell + m} }(1- \xi)^{ \al  + \ell +m    } (1+ \xi )^{  \beta + \ell + m}
\ee
These are smooth functions of $\theta, \phi$.   But $\theta, \phi$ are not a smooth coordinate patch   around the poles $\xi = \pm 1$.    We
need to express $ \cY^{\pm} _{n, l,m}$ as a smooth function on all of $U_{\pm} '$.
 This is accomplished when we pass to Cartesian coordinates as follows
\bigskip

\begin{lem}  \label{karlan} 
There exist   functions   $ \cY^{\pm} _{n, \ell,m}$
defined on a neighborhood of $ U'_{\pm}$ in $\bbR^3$  which are bounded and smooth with bounded derivatives and satisfy 
\begin{enumerate}
 \item
 $ \cY^{\pm} _{n,\ell,m}( x/ |x| )$
in spherical coordinates is $\cY^{\pm}_{n,\ell,m} (\xi , \phi) $ 
\item 
 $ \cY^{\pm} _{n,\ell,m}( x/ |x| )$ as  a function on $U_{\pm} $ in  $M = \bbR^3-\{0\}$ is in  $\cC^2(U_{\pm} ) $ and there is a constant $c$ such that   
 \be
| \pa_i  \B(   \cY^{\pm} _{n, l,m}(x/|x|)\B)|   \leq c|x|^{-1}   \hs    | \pa_i    \pa_j  \B(\cY^{\pm} _{n, l,m} (x/|x|) \B)| \leq    c|x|^{-2} 
 \ee 
 \end{enumerate}
     \end{lem} 
\bigskip

\pr  (1.) 
Consider  $ \cY^{+} _{n,\ell,m}$.  Since $\beta = \al + 2n$ we have 
\be
\begin{split}
  (1- \xi)^{\frac12 \al    } (1+ \xi  )^{\frac12 \beta} =  &   (1- \xi^2)^{\frac12 \al }  ( 1 + \xi )^n  
=    (\sin \theta)^{\al} (1+ \cos \theta )^n \\
\end{split}
\ee
and we also have 
$
  e^{i(m + n) \phi }  =   e^{-i \al \phi }    = ( \cos  \ \phi - i \sin \phi ) ^{\al } 
 $
Thus for  the product  
\be
\begin{split}
    (1- \xi)^{\frac12 \al    } (1+ \xi  )^{\frac12 \beta}   e^{i(m + n) \phi }   
=  & (\sin \theta \cos \phi - i \sin \theta \sin  \phi  )^{\al} (1+ \cos \theta )^n  \\
=  & \B( \frac{ x_1 - i x_2 }{|x| } \B)^{\al} \B(1+  \frac{x_3}{|x|}  \B)^n  \\
\end{split}
\ee
If $\al$ is positive the first factor is bounded, smooth, etc.  Even if $n$ is negative,    the same holds for the  second factor   on $U_+$  where $x_3 \geq -\de |x|$  .  This gives the result 
since  $P^{\al  ,\beta}_{ \ell + m } (\xi )  = P^{\al  ,\beta}_{ \ell + m } (x_3/|x| ) $ is a polynomial.
\bigskip

If $\al$ is negative we combine (\ref{elf1}), (\ref{elf2})  and use  $\al + \ell + m = \ell -n $ and $ \beta + \ell + m = \ell + n $  to    write
\be \label{elf3}
 \cY^{\pm} _{n,\ell,m}(\xi , \phi )   =  \const (1- \xi)^{-\frac12 \al    } (1+ \xi  )^{-\frac12 \beta} \frac{d^{\ell + m}} { d \xi^{\ell + m} }(1- \xi)^{\ell -n} (1+ \xi )^{ \ell + n} e^{i(m +  n) \phi } 
 \ee
Now we use $
  e^{i(m + n) \phi }  =   e^{-i \al \phi }    = ( \cos  \ \phi  + i \sin \phi ) ^{-\al } $ to write 
\be
\begin{split}
    (1- \xi)^{-\frac12 \al    } (1+ \xi  )^{-\frac12 \beta}   e^{i(m + n) \phi }    
=  & (\sin \theta \cos \phi +  i \sin \theta \sin  \phi  )^{- \al} (1+ \cos \theta )^{-n}   \\
=  & \B( \frac{ x_1 + i x_2 }{|x| } \B)^{- \al} \B(1+  \frac{x_3}{|x|}  \B)^{-n}   \\
\end{split}
\ee
which is bounded, smooth, etc.   Since  $\ell + m$,  $\ell -n$, $\ell +n$ are all non-negative  the derivatives of $(1- \xi)^{\ell -n} (1+ \xi )^{ \ell + n} $
 just give a polynomial in $\xi = x_3/|x|$ and so this factor satisfies the hypotheses  as well.  
 
 The analysis on $U_{-} $ is similar. 
\bigskip

\noindent
(2.)  For the derivatives we compute for example
\be
\pa_k \B(   \cY^{\pm} _{n,\ell,m}\B( \frac{x}{|x|}  \B)  \B) = \sum_j  \pa_j \cY^{\pm} _{n, \ell,m}\B( \frac{x}{|x|}  \B) \B[ \frac{|x|^2 \de_{jk} -x_j x_k}{|x|^3} \B] 
\ee
which is  bounded by $c|x|^{-1}$.  The second derivative is similar. This completes the proof. 
\bigskip

\rem
The construction shows that monopole harmonics may not be  polynomials in $x/|x|$, unlike the usual spherical harmonics. 
\bigskip

\subsection{self-adjointness} 

For self-adjointness we start in sperical coordinates. 
We introduce the dense domain in $\hat \cH = L^2(\bbR^+,r^2dr )  \otimes L^2(\tilde E, d \Om) $:
\be
\hat \cD  = \textrm{  finite sums of  } \psi \otimes   \cY_{n,\ell,m} \hs  \psi \in \cD_0 \hs \ell \geq |n|, |m| \leq \ell
\ee

\begin{lem} { \ } \label{sun}
\begin{enumerate}
\item $\hat H = \oplus_{\ell}  (  h_{\ell}   \otimes I ) $ maps $ \hat \cD$ to itself.
\item $\hat H$ is essentially self-adjoint on $\hat \cD$. 
\item The semi-group $e^{-\hat H t} $ defined with the self-adjoint closure satisfies
\be \label{commune} 
e^{- \hat H t}  = \bigoplus_{\ell =|n|}^{\infty} ( e^{-h_{\ell} t } \otimes I )
\ee
and maps $ \hat \cD$ to itself.
\end{enumerate}
\end{lem} 
\bigskip

\pr 
\begin{enumerate}
\item This is clear since $h_{\ell} $ preserves $\cD_0$. 
\item
 $h_{\ell} $ is essentially self-adjoint on $\cD_0 $ by lemma \ref{sanpedro}.    It follows that   it follow   $ h_{\ell}   \otimes I $ 
is essentially self-adjoint in $L^2(\bbR^+ )  \otimes \cK_{n,\ell} $ on the domain of finite sums of 
vectors $\psi \otimes \cY_{n,\ell,m}$  with $\psi \in \cD_0$  and  for fixed $n, \ell$ and $|m| \leq \ell$. 
Then   $\hat H = \oplus_{\ell}  (  h_{\ell}   \otimes I ) $
is essentially self-adjoint in $\hat \cH$  on  $\hat \cD$. 
\item  Both sides of (\ref{commune})  are continuous semi-groups.  Taking derivatives we see that the generators
agree on $\cD$ by part 1.  Since this is a domain of essential self-adjointness the generators are the same and hence the identity. 
The domain $\cD$ is preserved since $e^{-h_{\ell} t } $ preserves $\cD_0$. 
\end{enumerate} 
\bigskip 

We want to translate this to a statement about $H$ back in Cartesian coordinates
on the Hilbert space $\cH = L^2(E, dx) $.
The domain $ \hat  \cD $   becomes 
\be
 \cD  = 
\textrm{  finite sums of  } \psi (|x| )    \cY_{n,\ell,m} (x/|x| ) \hs  \psi \in \cD_0 \hs \ell \geq |n|, |m| \leq \ell
\ee

 \begin{lem}  \label{four}
   $ \cD \subset  \cC^2(E) $ and  for $\Psi \in \cD$  we have  the asymptotics
  \be
   \begin{split}
    |\Psi  (x)|  =  &   \begin{cases}  \cO( |x|^{\frac12( \sqrt5 -1)} )    & x \to 0 \\
    \cO( |x|^{-1} ) &  x \to \infty
     \end{cases}  \\  
      |\nabla_k \Psi  (x)| =  &   \begin{cases}   \cO( |x|^{\frac12( \sqrt5 -3)} )    & x \to 0 \\
    \cO( |x|^{-2} ) &  x \to \infty
    \end{cases}  \\  
  |\nabla_j \nabla_k \Psi  (x)|  =  &   \begin{cases}  \cO( |x|^{\frac12( \sqrt5 -5)} )    & x \to 0 \\
    \cO( |x|^{-3} ) &  x \to \infty
    \end{cases}  \\  
  \end{split} 
   \ee
    Furthermore  $\Psi, \nabla_k \Psi , \nabla_j \nabla_k  \Psi $ are all in $\cH = L^2(E, dx) $
  \end{lem} 
  \bigskip

 \pr   It suffices to consider $\Psi(x)   = \psi(|x|) \cY_{n, \ell, m}(x/|x|) $ with $\psi \in \cD_0$.  
Then $\psi(|x|) $ has the stated asymptotics by lemma \ref{pinsky} and $\mu \geq \sqrt5 /2$.  Also  $\cY_{n, \ell, m}(x/|x|) $
is bounded by lemma \ref{karlan}.   Hence the result.   Note that $\Psi$ is bounded. 
 
 For the derivatives it suffices to  
  prove the bounds  separately in $U_{\pm}$, say $U_+$.
 In the trivialization on $U_+$ we have 
 \be
 \begin{split} 
&\B(\nabla_k  (  \psi  \otimes \cY_{n, \ell, m})  \B) _+(x)  =  \B(  \pa_k - i A^+_k(x) \B)   \psi (|x|)   \cY^+_{n, \ell, m} (x|x|^{-1} )    \\
=&  
  \frac{x_k}{|x| } \psi' (|x| )  \cY^+_{n, \ell, m} (x|x|^{-1}) 
 +    \psi (|x|)  \pa_k \B(   \cY^+_{n, \ell, m} (x|x|^{-1}) \B)   - i A^+_k(x)     \psi (|x|)   \cY^+_{n, \ell, m} (x|x|^{-1} )  \\
\end{split} 
 \ee
The first term is bounded except for the $ \psi' (|x| ) $
which has the stated asymptotics by lemma \ref{pinsky}.  
For the second term  combine the asymptotics for  $ \psi (|x|)  $ with the bound 
$| \pa_k ( \cY^+_{n, \ell, m} (x|x|^{-1})) | \leq     \cO(|x|^{-1} ) $ from lemma \ref{karlan} 
to get the stated bounds.
For the last term $|A_k^+| \leq  c|x|^{-1}  $ which combined with the asymptotics for $\psi(|x|) $ gives the bound. 
The statement that $\nabla_k \Psi$ is in  $ L^2(E, dx)$ follows from the asymptotics

The second derivatives are handled in the same way.  
This completes the proof. 
\bigskip

In Cartesian coordinates lemma \ref{sun} becomes:

\begin{lem}   \label{six}
 For  $\Psi \in \cD \subset   L^2(E, dx) $, $H = -\frac12 \sum_k \nabla_k^2$  satisfies $(H \Psi )^{\wedge} = \hat H \hat \Psi$ and 
\begin{enumerate}
\item
$H $ maps $\cD$ to itself. 
\item   $H$ is essentially self-adjoint on $\cD$.
 \item   $e^{-Ht}$ defined with the self-adjoint closure maps $\cD$ to itself. 
  \end{enumerate}
 \end{lem} 
\bigskip

\pr 
Since $\cD \subset \cC^2(E) $ by lemma \ref{four}   the     statement $(H \Psi )^{\wedge} = \hat H \hat \Psi$ is just a repeat  of section \ref{defn}.   The rest is a translation of 
of the results for $\hat H$ in lemma \ref{sun}  by the unitary change of variables  operator. 
\bigskip

\section{Stochastic integrals} \label{bundle4}

  We  recall some definitions of stochastic integrals. General references are  \cite{IkWa81}, \cite{Oks03}, \cite{Sim79}.
First  let  $X_t $ be   Brownian motion in $\bbR$ starting at $X_0 =x$.  It is  a Gaussian process with mean $x$ and  covariance $\textrm{Cov} (X_tX_s) = \min(s,t)$.  Let $\cF_t$ be the $\sigma$-algebra generated by  $\{ X_s\}_{0 \leq s \leq t}$ and 
let $Y_t$ be a real valued  process  which is non-anticipating in the sense that $Y_t$ is $\cF_t$ measurable.

The  \textit{Ito integral}  is  defined by
\be 
Z_t   = \int_0^t  Y_s   dX_s    =   \lim_{n \to \infty} \sum_{k=0}^{[2^nt]-1} Y \B ( \frac{k}{2^n}  \B) \B) \B[  X\B( \frac{k+1}{2^n}\B)  - X\B( \frac{k}{2^n}\B)  \B]
\ee  
The limit exists  in $L^2$   and   satisfies
\be
E ^x(  |Z_t|^2  )  =   \int_0^t  E^x( | Y_s|^2 ) \  ds  
\ee
provided the right side is finite.  $Z_t$ has expectation zero (and is a martingale). 
The equation $Z_t   = \int_0^t  Y_s   dX_s $ is also written as the stochastic differential equation
\be
dZ_t  = Y_t dX_t 
\ee

Similarly we  define the \textit{Stratonovich integral}   by   
\be
Z_t  = \int_0^t  Y _s  \circ  dX_s    =   \lim_{n \to \infty}  \sum_{k=0}^{[2^nt]-1}\frac12\B[  Y\B( \frac{k}{2^n} \B)  + Y \B( \frac{k+1}{2^n}  \B) \B]\B[  X\B( \frac{k+1}{2^n}\B)  - X\B( \frac{k}{2^n}\B)  \B]
\ee  
which we write as
\be
\underline{d} Z_t =   Y _t  \circ  dX_t  
\ee
The Stratonovich integral obeys many of the usual rules of calculus.   In particular if $Z_t = f(X_t) $
\be  
  f(X_t) = f(X_0)  + \int_0^t f'(X_s) \circ dX_s
\ee
which we write as the chain rule
\be \label{chainrule} 
\ud( f(X_t) )   = f'(X_t) \circ dX_t 
\ee

The difference  between the two integrals    is  the quadratic integral 
\be
\begin{split}
  &  \int_0^t  Y _s  \circ  dX_s   -   \int_0^t  Y_s   dX_s  \\
&  =    \lim_{n \to \infty} \sum_{k=0}^{[2^nt]-1} \frac12\B[  Y \B( \frac{k+1}{2^n}\B)   -Y\B( \frac{k}{2^n} \B)  \B]\B[  X\B( \frac{k+1}{2^n}\B)  - X\B( \frac{k}{2^n}\B)  \B] \\
& \equiv    \frac12  \int_0^t  \blan dY_s, dX_s \bran  ds  \\ 
\end{split}
\ee
which we write as
\be \label{laugh} 
 Y _t  \circ  dX_t    =     Y_t   dX_t  
+  \frac12 \blan dY_t, dX_t \bran  dt
\ee

The quadratic term  vanishes if $Y_t$ has finite variation. 
Sometimes  the quadratic term can be simplified. For example  
if   $Y_t = \al(X_t)$ for a continuously differentiable  function $\al$ then the quadratic term is    $\frac12< d\al(X_t) , dX_t > =  \frac12 \al'(X_t) dt$  and  so
\be
 \int_0^t   \al (X_s) \circ  dX_s   -   \int_0^t   \al (X_t)   dX_s = \frac12 \int_0^t  \al'(X_s) ds
 \ee
\bigskip

Now let   $X_t = (X^1_t, X^2_t, X^3_t) $ be Brownian motion in $\bbR^3$ starting at $X_0=x$ consisting    of three independent  Brownian motions.      
So $X_t$ is a Gaussian process with mean  $x \in \bbR^3$    and covariance $\textrm{Cov} (X^i_t X^j_s) = \de _{ij}  \min(s,t)$. 
If the start point $X_0 =x  \neq 0$  then $X_t \neq 0$ with probability one.   Thus we can also regard $X_t $ as a process in  $M = \bbR^3 - \{ 0\}$ which we do 
hereafter.

 The Ito integral of a non-anticipating    $Y_t$  taking values in $\bbR^3$ is now  (summation convention) 
\be \label{a}
 Z_t   =    \int_0^t   (Y_s)_k   dX^k_s  
\ee 
and it exists if 
\be \label{b} 
 \int_0^t  E^x ( | Y_s |^2 ) \  ds  < \infty
\ee
Similarly there is a Stratonovich integral    
\be
  \int_0^t   (Y_{s} )_k \circ  dX^k_s
  \ee  
and they differ by a quadratic integral. 
\bigskip

The following result  covers the cases of interest to us.

\begin{lem}  \label{steady} 
Let $f$ be a continuous  vector valued   function on $M = \bbR^3- {0} $ satisfying 
\begin{enumerate}
\item  $|f(x)| $ is   bounded for  $|x| \geq 1$
\item  $|f(x)| \leq \cO (|x|^{-\al}) $ for some
$\al < \frac32$ and $ |x| \leq 1$. (So  $\int_{|x| \leq 1} |f(x)|^2 dx < \infty$.)
\end{enumerate}
  Then for Brownian motion starting at $x \in M$  and $t>0$
  \be \label{c} 
 \int_0^t  E^x( | f(X_s) |^2 ) \  ds  <  \infty
\ee
Hence condition (\ref{b})   is satisfied     and  the stochastic integral   $ \int_0^t   f_k(X_s)    dX^k_s$    exists. 
\end{lem} 
\bigskip

\pr  For $0 < s<t  $
\be
 E^x( | f(X_s) |^2 )  =  (2 \pi s)^{-\frac32} \int   | f(y) |^2  e^{- |x-y|^2 /2s }dy  
 \ee
For  $|y| \geq 1$,   $| f(y) |^2$ is bounded by a constant, the integral gives one and so the result is finite.
For $|y| \leq 1$ we use    $|f(y)|^2 \leq \cO (|y|^{-2\al}) $, then enlarge the integral to all space,  and have
a bound which is a constant times
\be 
   (2 \pi s)^{-\frac32} \int  |y|^{-2 \al }   e^{- |x-y|^2 /2s }dy   =  (2 \pi s)^{-\frac32} \int  |x-y |^{-2\al }   e^{- |y|^2 /2s } dy 
\ee
We break the integral into two regions. 
The first is  $|x-y| \geq |x|/2 $,  and in this  case the integral is bounded by $4|x|^{-2 \al } $.
The second is    $|x-y| \leq |x|/2 $.   In this case we  have $|y|  \geq |x|/2$ which implies   
$
 e^{- |y|^2 /2s } (2 \pi s)^{-\frac32}    \leq  \cO ( |x|^{-3} )
$.
Then the integral over the second region is bounded by  
\be
   \cO ( |x|^{-3} )  \int_{|x-y| \leq |x|/2 }   |x-y|^{-2\al }  dy =     \cO ( |x|^{-3} ) \int_0^{|x|/2}r^{2-2\al} dr     =  \cO( |x|^{- 2\al } ) 
 \ee
Hence everything is finite  and the proof is complete.
\bigskip

This result  still holds if  $f_k $ takes values in a  normed vector space   (for example $E_x$ )     rather than just $\bbR$.

 \section{Stochastic parallel translation} \label{bundle5}

For parallel translation along a Brownian path $X_t$ we  would like to define a stochastic parallel translation operator $\Pi_t = \Pi(t,0):  E_{X_0} \to E_{X_t} $.
This should involve 
 replacing    functions $ \exp ( i    \int_0^t  A^{\pm} _k(x_s)  dx^k_s ) $ in section \ref{bundle2}  by   stochastic  $ \exp ( i    \int_0^t  A^{\pm} _k(X_s) \circ  dX^k_s ) $ with   Stratonovich  integrals.  
 However we have to be mindful  of the fact that the Brownian path may visit each of  $U_{\pm}$ multiple times.   Our treatment roughly follows Norris \cite{Nor92}. 
 
 For Brownian paths  $X_t$  one  can find stopping times $\tau_0 =0 , \tau_1, \tau_2, \dots $ such that $\tau_k \to \infty$ as $k \to \infty$ and such that for  $t \in [ \tau_k, \tau_{k+1} ]$
 the $X_t$ is entirely contained in at least one of $U_{\pm}$. 
  For details see for example \cite{HaTh94}, p 426.

 Fix the specification of   stopping times and suppose we have defined $\Pi(t,0)  $ for $ t \in   [0, \tau_k ]$.   We extend the definition to     $ t \in   [0, \tau_{k+1} ]$
 defining 
 \be
 \Pi(t,0) =  \Pi(t,\tau_k) \Pi(\tau_k,0) 
 \ee
  If  $X_t \in   U_{\pm}$ for $ t \in   [\tau_k, \tau_{k+1} ]  $ and $v \in E_{X_{\tau_k}} $   then $\Pi(t, \tau_k) v \in E_{X_t} $ 
   is defined by  $( \Pi(t,\tau_k) v )_ {\pm} =  \Pi^{\pm}(t,\tau_k) v_{\pm}$ where
  \be  \label{sting}
\Pi^{\pm}(t,\tau_k) =    \exp \B( i    \int_{\tau_k}^{t }   A^{\pm} _k(X_s) \circ  dX^k_s \B)   
 \ee 
So our precise definition  of $\Pi_t:  E_{X_0} \to E_{X_t} $ is
\be \label{precise}
\Pi_t= \Pi(t,0) =  \sum_{k=0}^{\infty} 1( \tau_k < t \leq \tau_{k+1} )  \Pi(t,\tau_k) \Pi(\tau_k,0) 
\ee
The sum is actually finite due to the $\tau_k \to \infty$ condition.

\begin{lem}
$ \Pi(t,0) $ is well-defined since
\begin{enumerate}
\item  If  $X_t \in   U_{\pm}$ for $ t \in   [\tau_k, \tau_{k+1} ]$  the integral   $ \int_{\tau_k} ^t  A^{\pm} _k(X_s) \circ  dX^k_s$ exists.
\item   If  $X_t \in   U_+ \cap U_-  $ for $ t \in   [\tau_k, \tau_{k+1} ]  $ then  $\Pi^{\pm}(t,\tau_k) $ give the same $\Pi(t,\tau_k) $. 
\item  The definition is independent of the stopping times.  
 \end{enumerate}
 \end{lem} 
 \bigskip

 \pr 
 For the integral  suppose $X_t \in   U_+ $ for $ t \in   [\tau_k, \tau_{k+1} ]$ .  The quadratic term in this case   
can be 
 identified as the divergence $\textrm{div} A$ and we can  write the Stratonovich integral as  an Ito integral by 
\be
 \int_{\tau_k} ^t  A_k^+(X_s) \circ  dX^k _s   =    \int_{\tau_k} ^t A_k^+(X_s)   dX^k  _s   + \frac12  \int_{\tau_k} ^t  (\textrm{div} A^+) (X_s) ds
\ee
However  div$A^+ =0$  as one can see  from (\ref{walnut}).
  Therefore the 
 Stratonovich and Ito integrals coincide and we work with the Ito integral. 
 We  have
 \be
  \int_{\tau_k} ^t  A^+ _k(X_s)   dX^k_s  =    \int_0^t \bar  A^+  _k(X_s)   dX^k_s 
\ee
where
\be 
 \bar  A^+ _k(X_s) =    A^+  _k(X_s) 1( s \geq \tau_k ) 
\ee 
 From (\ref{itsy}) we have   $|A^+(x) | \leq  C|x|^{-1} $ for all  $x \in M$ so that  
 $
 | \bar  A^+ (X_s) | \leq   C |X_s|^{-1}    
 $.
The existence of the integral  then follows from
 \be \label{tuna}
 \int_0^t  E^x\B(  | \bar A^+ (X_s) |^2    \B)ds 
 \leq C     \int_0^t  E^x\B(   | X_ s|^{- 2}     \B)ds    <  \infty
\ee
where the second inequality follows from lemma \ref{steady} with $\al =1$. 
 The first point is established.

For the second point the independence of $\pm$ follows just as in the deterministic case thanks to the 
identity
\be
\int_{\tau_k}^t  ( \pa_k \phi) (X_s) \circ dX^k_s  = \phi(X_t) - \phi(X_{\tau_k})
\ee

Finally  we show that  the definition is independent of  the choice of stopping times $\tau_k$.  
Suppose we add an intermediate stopping time $\tau_k  < \si_k <  \tau_{k+1}$.   Then for 
$  t  \in [\tau_k,  \si_k] $ the definition is the same as before.  For  $t \in [  \si_k, \tau_{k+1}]$
the definition is now
\be
 \Pi(t,0) =  \Pi(t,\si_k) \Pi(\si_k, 0)  =  \Pi(t,\si_k) \Pi(\si_k, \tau_k) P(\tau_k,0)  
 \ee
and we want to compare this with the original  $ \Pi(t,0) =  \Pi(t, \tau_k) P(\tau_k,0)  $.  Thus we
need to show for $t \in [  \si_k, \tau_{k+1}]$
\be  \Pi(t,\si_k) \Pi(\si_k, \tau_k) =  \Pi(t, \tau_k)
\ee
 But  for $t \in [\tau_k, \tau_{k+1} ] $ we  are in some $U_{\pm} $ say $U_+$,   and so the identity follows by 
\be
\begin{split}
\Pi^+(t,\si_k) \Pi^+ (\si_k, \tau_k)  
 =  &   \exp \B( i    \int_{\si _k} ^t  A^{+} _k(X_s) \circ  dX^k_s \B)  \exp \B( i    \int_{\tau _k} ^{\si_k}  A^{+ } _k(X_s) \circ  dX^k_s \B)  \\
 =  &   \exp \B( i    \int_{\tau_k} ^t  A^{+} _k(X_s) \circ  dX^k_s \B)  \\
 =  &    \Pi^+(t, \tau_k)
\end{split}
\ee

   This result implies that if one choice of stopping times is a refinement of another then they
give the same result.   Finally any two choices of stopping times agree with their common refinement and hence give the same result.
This completes the proof. 
\bigskip

\begin{lem} 
\label{strat1} 
Let $\Psi \in \Ga(E)$ with bounded covariant  derivatives.       Then    $ \Pi_t^{-1}  \Psi(X_t)$ is a non-anticipating stochastic process taking values in $E_{X_0} = E_x$ and 
\begin{enumerate}
\item The Stratonovich differential satisfies
\be
\label{d1}
\ud \B( \Pi_t^{-1}  \Psi(X_t) \B) =    \Pi_t^{-1} (\nabla_k \Psi)(X_t) \circ dX^k_t
\ee  
\item The Ito  differential satisfies
\be
\label{d2}
d \B( \Pi_t^{-1}  \Psi(X_t) \B) =    \Pi_t^{-1} (\nabla_k \Psi)(X_t)  dX^k_t    - \Pi_t^{-1} (H \Psi ) (X_t)   
\ee  
\end{enumerate}
\end{lem} 
\bigskip

\pr  (1.)    In general if $s<t$ we define $\Pi(s.t) = \Pi(t,s)^{-1} $  Take stopping times $0 < \tau_1 < \tau_2  <  \dots   \to \infty$ as before. 
For $ t \in [\tau_k,\tau_{k+1}] $ 
 \be
\Pi_t^{-1} =  \Pi(t,0)^{-1}  =  \Pi(\tau_k,0) ^{-1}   \Pi(t,\tau_k)^{-1}   =   \Pi(0,\tau_k)   \Pi(\tau_k,t)   
 \ee
 Then our precise defintion is
 \be \label{precise1}
\Pi_t^{-1}  \Psi(X_t)     =  \sum_{k=0}^{\infty} 1( \tau_k < t \leq \tau_{k+1} ) \Pi(0,\tau_k)   \Pi(\tau_k,t)   \Psi(X_t) 
\ee
Then we have for the Stratonovich differential
\be \label{precise2} 
\ud \B( \Pi_t^{-1}  \Psi(X_t)\B)    =  \sum_{k=0}^{\infty} 1( \tau_k < t \leq \tau_{k+1} ) \Pi(0,\tau_k)  \ud  \B(   \Pi(\tau_k,t)   \Psi(X_t) \B) 
\ee
 Let us check this.  It means
 \be \label{santa}
 \Pi_t^{-1}  \Psi(X_t)  - \Psi(x)  =   \sum_{k=0}^{\infty}  \Pi(0,\tau_k)  \int_0^t 1( \tau_k < s \leq \tau_{k+1} ) \ud  \B(   \Pi(\tau_k,s)   \Psi(X_s) \B) 
 \ee
 Choose $n$ so  $t \in (\tau_n, \tau_{n+1}]$.   Then right side of (\ref{santa})  is
 \be
 \begin{split}
 & \sum_{k=0}^{n-1}  \Pi(0,\tau_k)  \int_{\tau_k}^{\tau_{k+1} } \ud  \B(   \Pi(\tau_k,s)   \Psi(X_s) \B) 
+ \Pi(0,\tau_{n})  \int_{\tau_{n}}^t  \ud  \B(   \Pi(\tau_n,s)   \Psi(X_s) \B) \\
& =  \sum_{k=0}^{n-1}  \Pi(0,\tau_k)   \B(  \Pi(\tau_k,\tau_{k+1})  \Psi(X_{\tau_{k+1}}) -  \Psi(X_{\tau_{k}})  \B) \\
&+ \Pi(0,\tau_{n})   \B(  \Pi(\tau_{n},t)  \Psi(X_{t}) -  \Psi(X_{\tau_{n}})  \B) \\
& =  \sum_{k=0}^{n-1}  \B(  \Pi(0,\tau_{k+1})  \Psi(X_{\tau_{k+1}}) -   \Pi(0,\tau_{k}) \Psi(X_{\tau_{k}})  \B) 
+  \B( \Pi(0,t)  \Psi(X_t) - \Pi(0, \tau_n) \Psi(X_{\tau_{n}})  \B) \\
&=   \Pi_t^{-1}  \Psi(X_t)  - \Psi(x) \\
\end{split}
\ee
 which is the  same as the left side of (\ref{santa}).    Thus (\ref{precise2}) is  confirmed.

Now   it suffices to show  that  restricted to the event  $\{ \tau_k < t \leq \tau_{k+1} \}$
\be
\ud \B(  \Pi(\tau_k,t)   
 \Psi(X_t) \B) =     \Pi(\tau_k,t)    (\nabla_j \Psi)(X_t) \circ dX^j_t
\ee  
which is an  identity in  $E_{X_{\tau(k)} }$.
This is sufficient since if we make this substitution in $(\ref{precise2})$ we get   $ \Pi_t^{-1} (\nabla_k \Psi)(X_t) \circ dX^k_t$.

By construction we have $X_t$ in at least one of $U_{ \pm} $ for $ \tau_k < t \leq \tau_{k+1}$.   Suppose it is $U_+$.
  In the trivialization on $U_+$ the  claim is that 
\be
\label{swift}
\ud \B(  \Pi^+(\tau_k,t)  
 \Psi^+(X_t) \B) =     \Pi^+(\tau_k,t)   \B(  (\pa_j -iA_j^+) \Psi_+ \B)(X_t) \circ dX^j_t
\ee  
Now 
\be  \Pi^+(\tau_k,t) =  \exp\B( -i \int _{\tau_k}^t  A^{\pm} _j(X_s) \circ  dX^j_s  \B)
\ee
The integral here is a semimartingale (in fact a martingale since it coincides with the Ito integral) 
and so  $\Pi^+(\tau_k,t)$ is a semimartingale. 
For semimartingales $Z_t$  we
have a chain rule $ \ud f(Z_t) = f'(Z_t) \ud Z_t$  \cite{IkWa81}.   Therefore
\be
\begin{split}
\ud \ \Pi^+(\tau_k,t) =&  \Pi^+(\tau_k,t)   \ud \B(- i \int^t_{\tau_k} A^{+} _j(X_s) \circ  dX^j_s   \B)  \\
 = &   \Pi^+(\tau_k,t)   \B(- i  A^{+ } _j(X_t)  \B)\circ  dX^j_t    \\
 \end{split} 
\ee
Furthermore   by the three dimensional version of (\ref{chainrule}) 
\be
\ud \    
 \Psi_+(X_t )  = \pa_j    \Psi_+(X_t ) \circ  dX^j_t       
\ee
Combining the last two with the product rule (for semimartingales)  gives   (\ref{swift}) and completes the proof of part (1.) .
\bigskip

(2.) We   change from a Stratonovich  differential to an Ito differential  as in   (\ref{laugh}) 
\be
\label{esq3}
       \Pi_t^{-1} (\nabla_k \Psi)(X_t) \circ dX^k_t
=        \Pi_t^{-1} (\nabla_k \Psi)(X_t)   dX^k_t   + \frac12  \blan  d  \B(  \Pi_t^{-1} (\nabla_k \Psi)(X_t)\B) , dX^k_t \bran  
\ee 
In the quadratic term  here we can exchange Stratonovich differentials and  Ito differentials since the difference
is the differential of a function with finite variation.    Then   by (\ref{d1}) again it is
\be \label{esq4}
\begin{split}
 \frac12  \blan  \ud  \B( \Pi_t^{-1} (\nabla_k \Psi)(X_t)\B) , dX^k_t \bran  
= & \frac12  \blan   \Pi_t^{-1} ( \nabla_j \nabla_k  \Psi )(X_t) \circ dX^j_t, dX^k _t \bran \\
= & \frac12  \blan   \Pi_t^{-1} (\nabla_j \nabla_k  \Psi )(X_t)  dX^j_t, dX^k_t \bran \\
= & \frac12    \Pi_t^{-1} (\nabla_k \nabla_k  \Psi )(X_t) dt   =   - \Pi_t^{-1} (H  \Psi )(X_t) dt  \\
\end{split} 
\ee  
Here we used   $   <dX^j_t, dX^k_t > = \de_{ij} dt $.  This completes the proof. 
\bigskip

This result would be adequate if for  example we took $\Psi \in \cC^{\infty} _0(E) $.  But we want $\Psi \in \cD$
which means  we have to  relax the condition that $\Psi$ have bounded covariant derivatives and allow growth as $x \to 0$. 

\begin{lem}  \label{strat2}
The results of lemma \ref{strat1} still hold if  $\Psi \in \cD$.
\end{lem} 
\bigskip

\pr     The expression  (\ref{d2}) for the Ito differential
means that 
\be
 \Pi_t^{-1}  \Psi(X_t)  =  \Psi(x)  + \int_0^t   \Pi_s^{-1} (\nabla_k \Psi)(X_s)  dX^k_s    - \int_0^t \Pi_s^{-1} (H \Psi ) (X_s)ds   
\ee  
We need to check that these integrals exist for $\Psi \in \cD$. 
 By  lemma \ref{four} we have $(\nabla_k  \Psi)  (t,x )$ is bounded for 
$|x| \geq 1$ and is $\cO( |x| ^{\frac12( \sqrt5 -3) } ) $ for  $|x| \leq 1$.  Hence the same holds for $\Pi_t^{-1} (\nabla_k  \Psi  )  (t,x )$.
The asymptotics as $|x| \to 0$ is less severe than $\cO(|x|^{-\frac32})$  so the existence of the first integral  follows by lemma \ref{steady}. 
For the second integral $H\Psi$ is again in $\cD$, hence it is  bounded and the integral exists.

\section{The Feynman- Kac formula } \label{bundle6}

We now prove the Feyman-Kac   formula, roughly following the strategy of Norris \cite{Nor92}.

\begin{thm} \label{one} 
Let $\Psi \in \cH = L^2(E,dx) $, let $X_t$ be Brownian motion with $X_0 =x$, and let $\Pi_t$ be
the stochastic   parallel translation operator.  Then   
\be
(e^{-Ht} \Psi )(x) =  E^x\B(  \Pi_t^{-1} \Psi  (X_t)  \B )
\ee
\end{thm}
\bigskip

\pr    It suffices to prove the result pointwise  for $\Psi$ in our  dense  domain $\cD$ of smooth sections.  
This is true since both sides of the equation define bounded operators on $L^2(E)$.  This is true for the left side since $H$ is a positive operator.
To see it is true for the right side note that   $| \Psi | \in L^2(M)=L^2(\bbR^3) $  so we can estimate  for $\Psi \in \cD$
\be
\begin{split}
&  \int  |E^x(   \Pi_t^{-1} \Psi  (X_t)  )|^2 dx    \leq   \int   \B( E^x  |\Psi  (X_t) | \B) ^2dx \\  & =  \int  | ( e^{- \De t }  | \Psi|  ) (x) |^2 dx = \|  e^{- \De t }  | \Psi |   \|^2
 \leq \| \Psi \|^2\\
\end{split} 
\ee
Hence the right side   extends to a bounded operator on all $L^2(E,dx) $.

Now with $\Psi \in \cD$ define  $\chi(t,x)$   for fixed $T$ and $t \leq T$ by 
\be
 \chi(t,x) =   (e^{-H(T-t) } \Psi )(x)
\ee 
Then $\chi(t, \cdot) $  is still in the domain $\cD$ by lemma \ref{six}.

Since   $\chi$ is a function of $t$ as well as $x$ the equation (\ref{d1}) becomes 
\be
\label{esq2}
\ud \B( \Pi_t^{-1}  \chi(t, X_t) \B) =  \Pi_t^{-1} \frac{\pa \chi }{\pa t} (t, X_t) dt  +      \Pi_t^{-1} (\nabla_k \chi)(t,X_t) \circ dX^k_t
\ee  
Changing to the Ito differential  by lemma \ref{strat2}  gives  
\be
d  \B( \Pi_t^{-1}  \chi(t, X_t) \B)  = \Pi_t^{-1} \frac{\pa \chi }{\pa t} (t, X_t)dt +  \Pi_t^{-1} (\nabla_k \chi)(t,X_t)   dX^k_t  - \Pi_t^{-1} (H  \chi )(t,X_t) dt
\ee
But  $ \pa \chi /\pa t  -  H \chi  =0$  
so simplifies  this simplifies to 
   \be
  \begin{split}
\label{esq5}
d \B( \Pi_t^{-1}  \chi(t, X_t) \B) 
=   &    \Pi_t^{-1} (\nabla_k \chi)(t,X_t)   dX^k_t   
 \\
\end{split} 
\ee   
which means  that
\be  \label{paul} 
  \Pi_t^{-1}\chi  (t, X_t)  =   \chi  (0, x)  +  \int_0^t   \Pi_s ^{-1}  (\nabla_k  \chi )  (s, X_s)    dX^k _s 
\ee
The  Ito integral in (\ref{paul})   has expectation zero.  Taking expectations  in this equation
gives $E^x(  \Pi_t^{-1}\chi  (t, X_t) )  =   \chi  (0, x)  $.    At $t=T$ this says
\be
E^x\B(  \Pi_T^{-1}\Psi(  X_T) \B)     =   (e^{- H T}\Psi)(x) 
\ee
Since $T$ is arbitrary this is our result.

\end{document}